\documentclass[useAMS,usenatbib]{mn2e}
\usepackage{amsmath}
\usepackage{graphicx,amssym,color}
\newcommand{\bc}{\begin{center}}
\newcommand{\ec}{\end{center}}

\title[AGN feedback and $\alpha$ enhancement in early-type galaxies]
      {AGN feedback and the origin of the $\alpha$ enhancement in early type
      galaxies - insights from the GAEA model}   
\author[G.~De Lucia et al.]
       {Gabriella De Lucia$^{1}$\thanks{Email: delucia@oats.inaf.it}, 
        Fabio Fontanot$^{1}$, 
        Michaela Hirschmann$^2$\\
        $^1$INAF - Astronomical Observatory of Trieste, via G.B. Tiepolo 11, 
        I-34143 Trieste, Italy\\
        $^2$Sorbonne Universites, UPMC-CNRS, UMR7095, Institut d'Astrophysique 
        de Paris, 75014, Paris, France}
\begin{document}

\pagerange{\pageref{firstpage}--\pageref{lastpage}} 
\pubyear{2016}

\maketitle

\label{firstpage}

\begin{abstract}
We take advantage of our recently published model for Galaxy Evolution and
Assembly (GAEA) to study the origin of the observed correlation between
[$\alpha$/Fe] and galaxy stellar mass. In particular, we analyse the role
of radio mode AGN feedback, that recent work has identified as a crucial
ingredient to reproduce observations. In GAEA, this process introduces the
observed trend of star formation histories extending over shorter time-scales
for more massive galaxies, but does not provide a sufficient condition to
reproduce the observed $\alpha$ enhancements of massive galaxies.  In the
framework of our model, this is possible only assuming that any residual star
formation is truncated for galaxies more massive than $10^{10.5}\,{\rm
M}_{\odot}$. This results, however, in even shorter star formation time-scales
for the most massive galaxies, that translate in total stellar metallicities
significantly lower than observed. Our results demonstrate that (i) trends of
[$\alpha$/Fe] ratios cannot be simply converted in relative time-scale
indicators, and (ii) AGN feedback cannot explain alone the positive correlation
between [$\alpha$/Fe] and galaxy mass/velocity dispersion. Reproducing
simultaneously the mass-metallicity relation and the $\alpha$ enhancements
observed poses a challenge for hierarchical models, unless more exotic
solutions are adopted such as metal-rich winds or a variable IMF.
\end{abstract}


\section{Introduction}
\label{sec:intro}

Metal abundances of early-type galaxies have long been used to constrain their
star formation histories. These studies typically take advantage of the ratio
between $\alpha$ elements (O, Mg, Si, S, Ca, and Ti) and iron, [$\alpha$/Fe], 
that is known to increase with galaxy stellar mass or velocity dispersion
\citep[e.g.][]{Trager_etal_2000,Kuntschner_etal_2001,Thomas_etal_2010}.
$\alpha$ elements are released mainly by type II supernovae (SNII), originating
from massive progenitors with relatively short life-times ($\sim$3-20~Myr). In
contrast, Fe-peak elements are mainly contributed from type Ia supernovae
(SNIa), whose progenitors have life-times ranging from $\sim$30~Myr up to
several Gyrs. Therefore, the [$\alpha$/Fe] ratio quantifies the
relative importance of SNII and SNIa, and is sensitive to different parameters
including the relative abundance of massive stars i.e. the slope of the initial
mass function (IMF), the delay time distribution (DTD) of SNIa, differential
losses of metals via galactic winds, and the time-scale of star formation
i.e. the time-scale over which a significant fraction of the stars in the final
system formed
\citep*{Tinsley_1979,Worthey_Faber_Gonzalez_1992,Matteucci_1994}.  It has
become, however, common practice to interpret variations of [$\alpha$/Fe] in
terms of different time-scales of star formation: in a closed-box system, a
relatively large [$\alpha$/Fe] ratio is obtained if most of the stars formed
before SNIa could enrich the gas with significant amounts of Fe. Lower values
are obtained if star formation continues for longer times, allowing the
progressive incorporation of Fe into the star forming material.

The $\alpha$ enhancement of massive elliptical galaxies has been
considered traditionally as a serious problem for the hierarchical merging
scenario, where massive galaxies are expected to assemble later than their lower
mass counter-parts, from progressive mergers/accretion of smaller
systems. \citet{Thomas_and_Kauffmann_1999} used star formation histories of
elliptical galaxies extracted from a hierarchical model, and showed that 
model predictions were in stark contrast with the observed trends exhibiting
a {\it decrease} of [Mg/Fe] with increasing luminosity. One limit of this work
was that it was based on a closed box approximation and did not model
chemical enrichment in a self-consistent way, i.e. by tracing the metal flows 
among different galaxy components. In addition, the adopted semi-analytic model
\citep{Kauffmann_and_Charlot_1998}, predicted bright ellipticals to be 
{\it younger} than their fainter counterparts, in contrast with observational 
evidence. 

The star formation histories predicted by the latest renditions of 
hierarchical models are in much better agreement with trends inferred from 
data. This success can be largely ascribed to the inclusion of `radio-mode' 
feedback, i.e. the suppression of late gas cooling (and hence star formation) 
in relatively massive haloes. \citet{DeLucia_etal_2006} showed that 
accounting for this process naturally gives rise to shorter time-scales 
of star formation for more massive ellipticals, in qualitative agreement with 
the conventional interpretation given to the observed increase of 
[$\alpha$/Fe] ratios
as a function of galaxy stellar mass. This work, however, was based on an 
instantaneous recycling approximation for the newly produced metals, and
therefore was not able to provide direct predictions for the [$\alpha$/Fe]
ratios. Parallel work by \citet{Nagashima_etal_2005} included a sophisticated
chemical enrichment treatment in a semi-analytic model, assuming a top-heavy IMF
during starbursts. This work did not discuss the star formation histories
predicted for elliptical galaxies, but showed that no simple modification of
the adopted model was able to invert the negative slope predicted for the 
galaxy velocity dispersion-[$\alpha$/Fe] relation.

In the last decade, detailed chemical enrichment models tracing individual
chemical abundances have been implemented both in in hydrodynamical simulations
and in semi-analytic models of galaxy formation. This has allowed the problem
of the $\alpha$ enhancements of elliptical galaxies to be revisited in greater
details. \citet{Pipino_etal_2009} showed that the inclusion of radio-mode
feedback in their semi-analytic model was able to invert the slope of the
galaxy stellar mass-[$\alpha$/Fe] relation for massive ellipticals.
The predicted slope was, however, still relatively shallow and the scatter too
large, particularly for intermediate mass galaxies. The role of `quasar mode'
feedback, associated with galaxy mergers, in reproducing the observed trends
was highlighted in
\citet{Calura_and_Menci_2011}, although this work also used an approach based
on a post-processing of star formation histories extracted from a hierarchical
model. Consistent results have been obtained more recently
by \citet{Taylor_and_Kobayashi_2015} and \citet{Segers_etal_2016} based on
hydrodynamical simulations, although the effect of the specific AGN
implementation used in the former study appears somewhat weaker and does not
reproduce the observed trend.

In contrast with these results, \citet{Arrigoni_etal_2010} found that, albeit
introducing the correct trend for the star formation histories of model
galaxies, AGN feedback (in this case both quasar and radio mode were included)
was not the key factor in reproducing the mass-[$\alpha$/Fe] relation. In their
model, a good agreement with data could be achieved by assuming a mildly
top-heavy IMF. More recently, \citet{Yates_etal_2013} have argued that a
positive (but shallower than observed) slope in the mass-[$\alpha$/Fe] relation
of local elliptical galaxies can be obtained using a DTD with $\leq 50$
per cent of SNIa exploding within $\sim 400$~Myr, or including metal-rich winds
that can drive light $\alpha$ elements directly into the circumgalactic medium.
Stronger slopes are obtained assuming a variable IMF (e.g. as a function
of the integrated star formation rate in the galaxy) as in
\citet{Gargiulo_etal_2015} and \citet{Fontanot_etal_2016}. These arise from the 
fact that massive galaxies tend to be associated with IMFs top-heavier (thus
characterized by larger contributions from SNII) than less massive
galaxies. The models presented in
\citet{Arrigoni_etal_2010}, \citet{Gargiulo_etal_2015}, and 
\citet{Fontanot_etal_2016} predict a flat mass-[$\alpha$/Fe] relation when 
assuming a universal IMF, although AGN feedback is included in all of them. 
This is in contrast with claims from e.g. \citet{Segers_etal_2016}. 
\section{The galaxy formation model}
\label{sec:simsam}

In this work, we use our recently published model for GAlaxy Evolution and
Assembly \citep*[GAEA -][]{Hirschmann_etal_2016}.  The model is based on that
described in \citet{DeLucia_and_Blaizot_2007}, with modifications introduced to
follow more accurately processes on the scales of the Milky Way
satellites \citep*{DeLucia_and_Helmi_2008,Li_etal_2010}. Our new model features
a sophisticated chemical enrichment scheme that accounts for the finite
lifetime of stars and non instantaneous recycling of metals, gas and energy
from asymptotic giant branch stars (AGBs), SNIa, and SNII \citep[for details,
see][]{DeLucia_etal_2014}. 
GAEA also features an updated stellar feedback scheme that adopts
parametrizations based on results from hydrodynamical simulations (the FIRE
feedback scheme defined in
\citealt{Hirschmann_etal_2016}). As shown in that paper, this model is able 
to reproduce both the evolution of the galaxy stellar mass function and that
observed for the correlation between galaxy stellar mass and gaseous
metallicity.

Our galaxy formation model is coupled to the Millennium Simulation 
\citep{Springel_etal_2005}, a high resolution cosmological simulation 
following 2,160$^3$ particles of mass $8.6\times10^8\,{\rm M}_\odot$, from
$z=127$ to present, in a cubic region of $500\,{\rm Mpc}\,{\rm h}^{-1}$
comoving on a side, and based on a WMAP1 cosmology ($\Omega_{m}=0.25$,
$\Omega_{b}=0.045$, $\Omega_{\lambda}=0.75$, $h=0.73$, and
$\sigma_8=0.9$). More recent measurements of cosmological parameters provide
slightly different values and, in particular, a larger value for $\Omega_{m}$
and a lower one for $\sigma_8$. We do not expect our predictions to be
significantly affected by these differences
\citep[see e.g.][]{Wang_etal_2008}.  For the following analysis, we have used 
only about 10 per cent of the entire volume of the Millennium Simulation, but 
we have verified that results would not change when using a larger volume.

\section{$\alpha$ enhancement of model galaxies}
\label{sec:enhancement}

\begin{figure}
\centering
\resizebox{8.7cm}{!}{\includegraphics{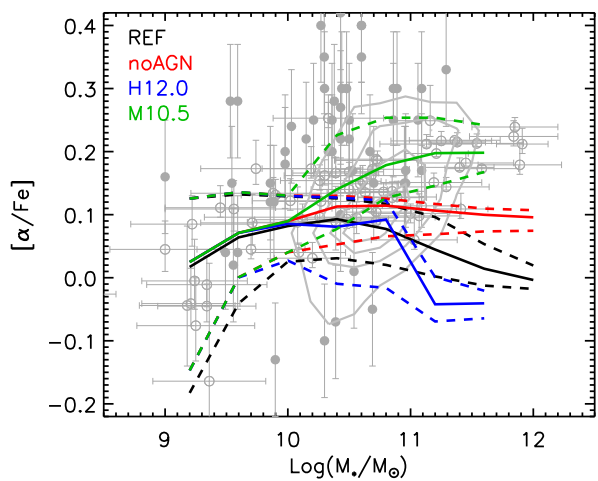}} 
\caption{The relation between [$\alpha$/Fe] and galaxy stellar mass. Grey
symbols show observational data from \citet[][filled]{Spolaor_etal_2010},
and \citet[][empty]{Arrigoni_etal_2010}. The grey contours show observational
estimates by \citet{Thomas_etal_2010}. The black and red solid lines show the
median relations corresponding to our reference run and a run where AGN
feedback has been switched off, respectively. The blue and green lines
correspond to two different toy models where either gas cooling is suppressed
when halo mass becomes larger than $10^{12}\,{\rm M}_{\odot}$, or star
formation is truncated when the galaxy stellar mass becomes larger than
$10^{10.5}\,{\rm M}_{\odot}$. Dashed lines show the 10th and 90th percentiles
of the distributions.
\label{fig:alphamass}}
\end{figure}

Fig.~\ref{fig:alphamass} shows the relation between [$\alpha$/Fe]\footnote{For
model galaxies, this is approximated by [O/Fe], as oxygen is the most abundant
of the $\alpha$ elements.} and galaxy stellar mass as predicted by our
reference run (REF, black lines), and by additional runs that we discuss
below. Grey symbols with error bars correspond to observational estimates
by \citet[][filled symbols]{Spolaor_etal_2010} and \citet[][empty
symbols]{Arrigoni_etal_2010}. The grey contours show measurements
by \citet{Thomas_etal_2010}. Only model galaxies with a bulge-to-total stellar
mass ratio larger than 0.7 (results are not very sensitive to this cut) have
been considered. Solid lines correspond to the median of the distributions,
while dashed lines show the 10th and 90th percentiles. Our REF run predicts a
relatively flat relation, with a bend towards {\it lower} [$\alpha$/Fe] ratios
for galaxies more massive than $\sim 10^{11}\,{\rm M}_{\odot}$. This is in
contrast with observational measurements, that show an enhancement of
[$\alpha$/Fe] for the most massive galaxies. The red lines correspond to our
reference model with radio mode feedback switched {\it off} (noAGN). The
inclusion of radio-mode feedback, in our model, turns a relatively flat
relation into a relation with a {\it negative} slope. This is in contrast with
results from previous studies, and appears counter-intuitive as AGN feedback
does introduce the expected trend for the star formation histories (see
below). The blue and green lines correspond to two toy models where our default
AGN feedback model is off but that assume either (i) gas cooling is suppressed
when halo mass becomes larger than $10^{12}\,{\rm M}_{\odot}$ (blue lines in
the figure, H12.0) or (ii) star formation is truncated when the galaxy stellar
mass becomes larger than $10^{10.5}\,{\rm M}_{\odot}$ (green lines, M10.5). The
former corresponds to the implementation adopted in \citet{Pipino_etal_2009}.
In this case, the average mass-[$\alpha$/Fe] relation is close to that obtained
within our REF model, but there is an even more pronounced turn over at the
most massive end. The other toy model is the only one, among those we tested,
that predicts a positive slope.

\begin{figure*}
\centering
\resizebox{18cm}{!}{\includegraphics{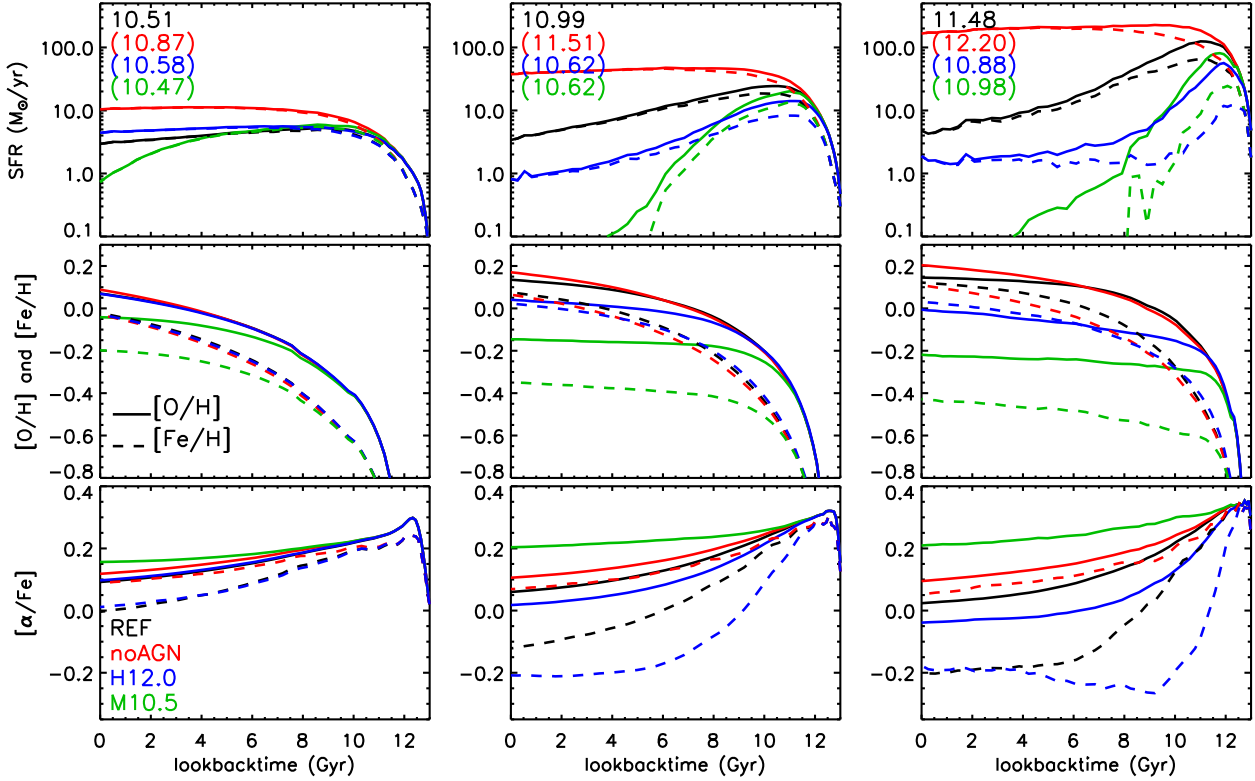}} 
\caption{Star formation histories (top panels), and evolution of metal
abundances (middle panels) and [$\alpha$/Fe] (bottom panels) for central
galaxies with log$({\rm M}_{\rm star}/{\rm M}_{\odot}) \sim 10.5$, $11$, and
$11.5$ in the reference run (from left to right). In the top panels, solid and
dashed lines correspond to the star formation histories obtained summing up the
contribution from all progenitors, and considering only the main progenitors,
respectively.  The legend indicate the average galaxy stellar mass at z=0 for
each run. In the middle panels, solid and dashed lines correspond to
the O and Fe abundances, respectively. In the bottom panels, solid and dashed
lines correspond to the [$\alpha$/Fe] of the stellar and gaseous components,
respectively. A green dashed line is not plotted because in this toy model gas
is accumulated but never used once log$({\rm M}_{\rm star}/{\rm M}_{\odot}) >
10.5$. In the middle and bottom panels, the evolution corresponds to that of
the main progenitor.
\label{fig:galhist}}
\end{figure*}

Fig.~\ref{fig:galhist} shows the average star formation histories (top row),
[O/H] and [Fe/H] abundances (middle row, solid and dashed lines respectively)
and [$\alpha$/Fe] (bottom row, solid for the stellar component and dashed for
the gas) for central galaxies with log$({\rm M}_{\rm star}/{\rm
M}_{\odot}) \sim 10.5$, $11$, and $11.5$ (from left to right) in our REF
run. Coloured lines correspond to the averages computed using the same galaxies
in the other runs, to better highlight the effect of the different
assumptions. This implies that the mean final mass of the galaxies in each
panel is different with respect to the REF run (as indicated in the legend). In
the top panels, solid lines correspond to the star formation histories computed
summing up the contributions from all progenitors, while dashed lines are
obtained considering only the main progenitor (the most massive at each
redshift).  The evolution of [O/H], [Fe/H], and [$\alpha$/Fe] plotted
corresponds to that of the main progenitor.  In our REF model, the most massive
galaxies form their stars on shorter time intervals than their less massive
counterparts. The final average [$\alpha$/Fe] ratios for the stellar mass bins
considered are, however, very similar. Switching off radio mode feedback
(noAGN) translates in a larger [O/H] value because more stars are formed at
late times, when the gas has been enriched to higher levels by previous
generations of stars. The [Fe/H] is slightly lower in the noAGN run because of
its delayed release by SNIa. The final values of the [$\alpha$/Fe] ratios in
the noAGN run are larger than in the REF run, but again very similar for the
three mass bins considered. Since the effect described is stronger for the most
massive galaxies, the predicted mass-[$\alpha$/Fe] relation is flatter than in
the REF run.

The H12.0 model predicts star formation histories that are in between the REF
and noAGN run for the lowest mass bin considered.  For the other two bins, the
shape of the star formation histories is close to that predicted by the REF run
but the overall levels of star formation are reduced. The almost systematic
difference between the predicted star formation rates in the last several Gyrs
translates in systematically lower abundances of both [O/H] and [Fe/H].
However, since a larger fraction of the stars in the final system originate
from lower abundance gas, the [$\alpha$/Fe] ratio decreases with respect to the
REF run.  Finally, in the M10.5 model, the star formation histories are
artificially {\it truncated} at increasing cosmic epochs for increasing stellar
mass. The [O/H] and [Fe/H] abundances (and therefore [$\alpha$/Fe]) of the
stellar component are effectively {\it frozen} at the times corresponding to
the last star formation events, and only vary because of mergers/accretion
events. The M10.5 model is the only one, among those considered, predicting
both the right trend for the star formation time-scales as a function of galaxy
stellar mass, and a positive slope of the mass-[$\alpha$/Fe] relation.

\begin{figure}
\centering
\resizebox{8.7cm}{!}{\includegraphics{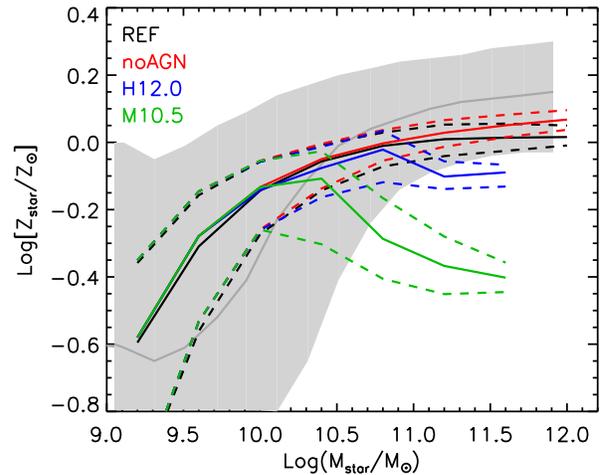}} 
\caption{Relation between the galaxy total stellar metallicity and galaxy
stellar mass at $z=0$. The grey region shows measurements
by \citet{Gallazzi_etal_2005}, based on the SDSS.
Coloured lines correspond to the different models used in this study.  Colours
and linestyles have the same meaning adopted in
Fig.~\ref{fig:alphamass}.\label{fig:massmet}}
\end{figure}

The time-interval over which most of the stars in the galaxies form has
important consequences also on their total metallicity. This can be appreciated
in Fig.~\ref{fig:massmet}, that shows the relation between galaxy stellar
metallicity and stellar mass at $z=0$. The grey region shows measurements
by \citet{Gallazzi_etal_2005}, based on the Sloan Digital Sky Survey
(SDSS). Solid coloured lines show the median relations obtained for our
different models. Predictions from our REF and noAGN runs are in quite good
agreement with data. The median metallicity of model galaxies is, in both runs,
above (below) the median of the observational data for stellar masses below
(above) log$({\rm M}_{\rm star}/{\rm M}_{\odot}) \sim 10.5$, but always within
the uncertainties. The two toy models considered in our study exhibit a bend of
the mass-metallicity relation towards lower metallicity for more massive
galaxies. This is particularly pronounced for the M10.5 model. This problem has
been discussed earlier in \citet{DeLucia_and_Borgani_2012}: in hierarchical
models, massive galaxies grow primarily by accretions of low mass galaxies:
stars form in different progenitors, most of them at high redshift, and then
assemble at later time. These accretion events typically involve very little
gas, and so negligible amounts of new stars form during merger driven
starbursts. Since the stars form at high redshift, from material that is
metal-poor, the final object also has low metallicity.  Late mergers and
accretions would tend to lower further the stellar metallicity. In reality,
some of the accreted low mass galaxies could be destroyed by efficient tidal
stripping. In addition, simulations show that stars accreted during minor
mergers tend to reside in the outer regions of the remnant galaxy
\citep{Naab_etal_2009,Hirschmann_etal_2015}, outside the region typically
considered for the observational estimates (a 3 arcsec diameter fibre for 
SDSS). Even neglecting, however, all stars accreted below $z\sim 1$ does not
alleviate the problem, simply driven by the fact that stars formed over
a too short time interval preventing a sufficient enrichment of the star
forming material.

\section{Discussion and Conclusions}
\label{sec:discconcl}

We have studied the origin of the relation between the [$\alpha$/Fe] ratio and 
galaxy stellar mass in the framework of our recently published model for 
Galaxy Evolution and Assembly 
\citep[GAEA,][]{DeLucia_etal_2014,Hirschmann_etal_2016}. In particular, we have 
focused on the role of AGN radio mode feedback that recent studies have
identified as a crucial ingredient to reproduce the observed trend as a
function of galaxy stellar mass \citep[][see
also \citealt{Pipino_etal_2009}]{Taylor_and_Kobayashi_2015,Segers_etal_2016}.
In the framework of our model, radio mode feedback introduces the observed
(inferred) trend of star formation histories extended on shorter time intervals
for more massive galaxies. Contrary to conventional expectations, however, this
is not a sufficient condition to reproduce the observed increase of the
[$\alpha$/Fe] ratio with increasing galaxy stellar mass. Switching off AGN
feedback destroys any correlation between galaxy stellar mass and star
formation time-scale (as expected), but has the consequence of flattening a
relation that is otherwise weakly negative. This is in contrast with recent
claims, and with the conventional interpretation of the [$\alpha$/Fe]-mass
relation. As discussed in \citet{Fontanot_etal_2016}, we have verified that no
simple modification of our chemical model (e.g. a different DTD for SNIa) is
sufficient to invert the predicted slope of the mass-[$\alpha$/Fe] relation. A
positive slope is obtained in GAEA when using a variable
IMF \citep{Fontanot_etal_2016}, or a toy model where any residual star
formation is truncated for galaxies more massive than $10^{10.5}\,{\rm
M}_{\odot}$ (our M10.5 model). For galaxies with log$({\rm M}_{\rm star}/{\rm
M}_{\odot}) \sim 11.5$, this means no stars are formed, on average, in the last
$\sim$~8~Gyrs. These very short time-scales of star formation (yet
significantly larger than those inferred by e.g. \citealt{Thomas_etal_2010})
make it very difficult for the most massive galaxies in this model
to reach the observed levels of total stellar metallicity.

Our results are only in apparent contradiction with previous studies. 
\citet{Pipino_etal_2009} were the first to reproduce a positive slope in the 
mass-[$\alpha$/Fe] relation in the framework of a hierarchical model.  They
argued that the success of their model was due to the inclusion of AGN radio
mode feedback. In fact, their model corresponds to one of the toy models used
in our study: gas cooling is suppressed in haloes more massive than
$10^{12}\,{\rm M}_{\odot}$ (our H12.0 model). In the framework of their model,
however, this parametrization gives results that are similar to those from our
M10.5 model: star formation is effectively suppressed in massive galaxies in
the last 8-9~Gyrs (see their Fig.~5). \citet{Pipino_etal_2009} also showed (see
their Fig.~8) that their model predicts total stellar metallicities for massive
galaxies that are significantly lower than observed, in agreement with results
shown above. Our results are also fully consistent
with \citet{Arrigoni_etal_2010}: their model including both quasar and radio
mode AGN feedback predicts a relatively flat [$\alpha$/Fe]-mass relation when
coupled with a chemical model assuming `standard' yields and IMF. The behaviour
of the three example galaxies discussed in \citet{Taylor_and_Kobayashi_2015} is
very close to that of our model including/excluding radio mode feedback (see
the blue arrows in their Figs.~11 and 13). \citet{Gargiulo_etal_2015} use a
semi-analytic model that includes AGN feedback prescriptions corresponding to
those of our reference run. They claim that a variable IMF is needed to
reproduce the observed $\alpha$ enhancements trends, as we also find for
GAEA \citep{Fontanot_etal_2016}. They also show that, in their model, the
distributions of the star formation time-scales explain the scatter in the
observed mass-[$\alpha$/Fe] relation, but are not responsible for the positive
slope. Their model galaxies, however, do not exhibit a trend for shorter
time-scales of star formation for more massive galaxies (see their
Fig.~14). \citet{Segers_etal_2016} use simulations from the Evolution and
Assembly of GaLaxies and their Environments (EAGLE) project and argue that AGN
feedback can account for the $\alpha$-enhancement of massive galaxies.  Also in
this case, the local mass-metallicity relation bends at the most massive end,
as shown in the recent work by \citet[][their
Fig.~4]{Okamoto_etal_2016}. Results from the hydrodynamical simulations
presented in this latter paper also support our conclusions and highlight the
difficulties in reproducing the $\alpha$ enhancements and mass-metallicity at
the same time.

Our study demonstrates that, generally speaking: (i) [$\alpha$/Fe] ratios
cannot be used as simple indicators of the star formation time-scale, and (ii)
AGN feedback is not the key ingredient to explain the $\alpha$ enhancement of
massive elliptical galaxies. Indeed, Fig.~\ref{fig:galhist} shows that three
out of the four runs used in this study predict shorter star formation
time-scales for more massive galaxies. In two of these models, the relation
between stellar mass and [$\alpha$/Fe] has a negative slope. In the framework
of our model, radio mode feedback would be able to introduce a positive slope
for the mass-[$\alpha$/Fe] relation only if effective in {\it truncating} star
formation in massive galaxies in the past several Gyrs.  This, however, would
translate in too low total stellar metallicities for massive
galaxies. Reproducing simultaneously the measured $\alpha$ enhancements and the
observed mass-metallicity relation in the local Universe represents a challenge
for hierarchical models of galaxy formation, unless more extreme solutions like
a variable IMF or $\alpha$ enhanced winds are invoked.

\section*{Acknowledgements}
We thank the referee for constructive comments that improved significantly the
clarity of our work. GDL and FF acknowledge financial support from the MERAC
foundation, and from the Italian Ministry of University and Research under the
contract PRIN-MIUR-2012 `The Intergalactic Medium as a probe of the growth of
cosmic structures'. MH acknowledges financial support from the European
Research Council via an Advanced Grant under grant agreement n. 321323
(NEOGAL).

\bsp

\label{lastpage}

\bibliographystyle{mn2e}
\bibliography{alphaen}

\end{document}